\documentclass[aps,prd,amsmath,amssymb,superscriptaddress,nofootinbib]{revtex4-2}

\usepackage{graphicx} % Include figure files
\usepackage{dcolumn}  % Align table columns on decimal point
\usepackage{bm}       % Bold math
\usepackage{physics}  % Physics notation package
\usepackage{hyperref} % Add hypertext capabilities
\usepackage{booktabs}
\begin{document}

\title{Factorization vs. Non-Factorization: S-Matrix Corrections for Precision Neutrino Physics}

\author{David Delepine}
\email{delepine@ugto.mx}
\affiliation{Physics Department, Divisi\'on de Ciencias e Ingenier\'ias, Universidad de Guanajuato, C.P. 37150, Le\'on, Guanajuato, M\'exico.}

\author{A. Yebra}
\email{azarael@fisica.ugto.mx}
\affiliation{Physics Department, Divisi\'on de Ciencias e Ingenier\'ias, Universidad de Guanajuato, C.P. 37150, Le\'on, Guanajuato, M\'exico.}

\date{\today}

\begin{abstract}
 The standard treatment of neutrino oscillations  usually  relies on  factorization which assumes neutrino production, propagation, and detection are independent processes. As a consequence, the total probability is given by the product of production, oscillation and detection probabilities.  As next-generation experiments are bringing neutrino physics to a high level of precision,  the validity of this assumption must be checked. We present an  S-matrix treatment of the entire experimental chain—pion decay, neutrino propagation, and nucleon interaction—as a single, coherent quantum process. Our results reveal non-factorizable terms arising from spin and angular correlations between production and detection final states. In the $\Delta L=0$ channel, these corrections introduce a $\sim 1\%$ systematic shift in the energy spectrum and a non-vanishing azimuthal asymmetry, important to be taken into account  for precision measurements of $\delta_{CP}$. For the $\Delta L=2$ Majorana channel, we demonstrate that the S-matrix formalism is generating  an azimuthal modulation that provides a direct way to access to the Majorana CP phases, which remain hidden in standard factorized effective mass approximations. 

\end{abstract}

\maketitle

%\section{Introduction}
%%%%%%%%%%%%%%%%%%%%%%%%%%%%%%%%%%%%
\section{Introduction}
%%%%%%%%%%%%%%%%%%%%%%%%%%%%%%%%%%%%%
In 1998, neutrino oscillations were discovered \cite{Super-Kamiokande:1998kpq, SNO:2001kpt}. This discovery implied that neutrinos are massive and mix through the  Pontecorvo-Maki-Nakagawa-Sakata (PMNS) lepton mixing matrix. More than 20 years later, neutrino physics is entering a new era of precision measurements of the PMNS mixing matrix elements. Hence, it is highly relevant to check if generally assumed hypotheses are valid or not. Traditionally, the experimental analysis of these oscillations is based on factorization, where the production of a neutrino (e.g., from pion decay), its subsequent oscillation over a baseline $L$, and its eventual detection via nucleon scattering are treated as three independent events \cite{Beuthe:2001rc}. Under this paradigm, the neutrino loses the ``angular memory'' of its birth before it reaches the detector \cite{Giunti:2002xg, Akhmedov:2009rb}. If the experimental chain is instead treated as a single, coherent quantum mechanical transition using an S-matrix approach \cite{Giunti:1993se, Grimus:1996av}, the neutrino acts as a virtual quantum link between production and detection. This coherence preserves spin and angular correlations between the source muon and the detection muon that are neglected in the standard factorized hypothesis.

In this paper, we compute the full transition amplitude $|T_{fi}|^2$ for both lepton-number-conserving ($\Delta L=0$) and lepton-number-violating ($\Delta L=2$) \cite{Bilenky:1987ty} processes. By isolating the non-factorizable terms, we identify Longitudinal Correlations, which manifest as scalar memory in the energy spectrum, and Transverse Correlations, which appear as a parity-violating azimuthal asymmetry \cite{LlewellynSmith:1971uhs}.

We show that these effects are numerically small—on the order of $1\%$—but they carry profound physical implications. For standard oscillations, accounting for these terms is necessary to avoid systematic biases in the measurement of the CP-violating phase $\delta_{CP}$ and the mass hierarchy \cite{DUNE:2020jqi}. In the search for Majorana neutrinos, the S-matrix formalism reveals that the Majorana CP phases shift the position of oscillation peaks and modulate the event rate as a function of the azimuthal angle $\phi$ \cite{Bilenky:1987ty}.

This paper is organized as follows: In Section II, we define  how to identify the  factorizable or non-factorizable terms; Section III develops the S-matrix formalism for $\Delta L=0$ processes and isolates the non-factorizable corrections; Section IV extends this treatment to $\Delta L=2$ Majorana conversions; Section V evaluates the experimental sensitivity to the next generation of neutrino experiments; and Section VI provides our concluding remarks on the transition from factorized approximations to full quantum coherence.

%%%%%%%%%%%%%%%%%%%%%%%%%%%%%%%%%%%%%
\section{ Factorizable vs. Non-factorizable}
%%%%%%%%%%%%%%%%%%%%%%%%%%%%%%%%%%%%%%%
The question of ``factorizability'' is fundamentally about whether we can treat the production of a neutrino and its subsequent detection as two independent events, or if they must be treated as one single, coherent quantum mechanical process \cite{Beuthe:2001rc}. In the standard (factorized) approach, we assume the neutrino is produced, loses memory of its birth, and then interacts \cite{Akhmedov:2009rb}.
In the S-Matrix Approach, the neutrino is a virtual particle that maintains spin correlations between the source and the detector \cite{Giunti:1993se}.

A process $\pi^+(p_1) \to \mu^+(p_2) + \nu \to \mu^-(p_l) + N'$ is defined as Factorizable if and only if its differential cross-section can be written as the incoherent product of the production probability, the propagation probability, and the detection cross-section \cite{Grimus:1996av}:
$$d\sigma \propto \underbrace{|\mathcal{M}_{prod}|^2}_{\text{Production}} \times \underbrace{P(\nu_\mu \to \nu_\mu)}_{\text{Oscillation}} \times \underbrace{|\mathcal{M}_{det}|^2}_{\text{Detection}}$$
It implies that no angular or spin correlation exists between the production muon ($p_2$) and the detection muon ($p_l$). Mathematically, it requires that all terms involving the direct contraction $(p_l \cdot p_2)$ or $\epsilon(p_l, p_2, \dots)$ must vanish or average to zero.

To prove whether the process is factorizable, we calculate the full squared amplitude $|T_{fi}|^2$ treating the neutrino as an internal propagator \cite{Giunti:1993se}, and we look for non-factorizable terms \cite{Grimus:1996av}. The process is non-factorizable if the squared amplitude contains terms that explicitly couple the kinematic variables of the source ($p_2$) to the kinematic variables of the detector ($p_l$) \cite{Beuthe:2001rc}. Two types of correlation between production and detection can be produced:
\begin{itemize}
\item Longitudinal Correlation: terms proportional to $\mathcal{C}_{scalar} (p_l \cdot p_2)$ \cite{LlewellynSmith:1971uhs}.
\item Transverse Correlation: terms proportional to the epsilon contraction $\mathcal{C}_{parity} \, \epsilon_{\mu\nu\rho\sigma} p_l^\mu p_2^\nu P^\rho q^\sigma$ \cite{LlewellynSmith:1971uhs}. The detector muon's azimuthal emission angle is correlated with the source muon's production plane. This term is generated by Standard Model (SM) parity violation interaction.
\end{itemize}

%%%%%%%%%%%%%%%%%%%%%%%%%%%%%%%%%%%%%%%%%%%%%%%%%%%%%%
\section{The S-Matrix Formalism for $\Delta L=0$}
%%%%%%%%%%%%%%%%%%%%%%%%%%%%%%%%%%%%%%%%%%%%%%%%%%%%%%
We treat the entire experimental chain—pion decay, neutrino propagation, and nucleon interaction—as a single coherent quantum process \cite{Giunti:1993se, Grimus:1996av}. First let us apply this formalism to a $\Delta L=0$ process. The total transition amplitude $T_{fi}$ is calculated for the process:
\begin{equation}
    \pi^+(p_1) \to \mu^+(p_2) + \underbrace{\nu_\mu \to \nu_{\mu}}_{propagator} + N(P) \to \mu^-(p_l) + N'(P')
\end{equation}
Using the Feynman rules for the V-A weak interaction \cite{LlewellynSmith:1971uhs}, the amplitude is:
\begin{equation}
T_{fi} = \left(\frac{G_F}{\sqrt{2}}\right)^2 V_{ud}^2 f_\pi \bar{u}(p_l) \Gamma_L S_F(q) \Gamma_P v(p_2) \cdot J_{had}^\mu
\end{equation}
where 
\begin{itemize}
    \item $v(p_2)$: This is the antiparticle spinor for the $\mu^+$ produced at the source (from $\pi^+ \to \mu^+ \nu$).
    \item $u(p_l)$: This is the particle spinor for the $\mu^-$ detected at the far end (from $\nu N \to \mu^- N'$) in case of $\Delta L=0$ process.
    \item $\Gamma_P$ (Production Vertex):This term comes from the pion decay current. Since the pion is spin-0, its hadronic current is just proportional to its momentum $p_\pi$. The effective vertex used in our formula absorbs the pion momentum $\not{p}_\pi$ to couple the lepton line:$$\Gamma_P = \not{p}_\pi (1 - \gamma_5)$$
    \item $\Gamma_L$ (Lepton Vertex at Detector): This is the standard Dirac matrix structure for the weak current at the detector side:$$\Gamma_L = \gamma_\mu (1 - \gamma_5)$$
    \item For Quasi-Elastic Scattering, the hadronic current is given by \cite{LlewellynSmith:1971uhs}:
    \begin{equation}
    J_{had}^\mu = \bar{u}_p(P') \left[ \underbrace{\gamma^\mu F_1(Q^2) + i\frac{\sigma^{\mu\nu}q_\nu}{2M} F_2(Q^2)}_{\text{Vector (Standard)}} + \underbrace{\gamma^\mu \gamma_5 F_A(Q^2) + \frac{q^\mu \gamma_5}{2M} F_P(Q^2)}_{\text{Axial (Spin-Dependent)}} \right] u_n(P)
    \end{equation}
    where:
    \begin{itemize}
        \item $F_1(Q^2)$: Dirac Form Factor (Charge distribution).
        \item $F_2(Q^2)$: Pauli Form Factor (Anomalous magnetic moment).
        \item $F_A(Q^2)$: Axial Form Factor. 
        \item $F_P(Q^2)$: Pseudoscalar Form Factor.
    \end{itemize}
\end{itemize}
$S_F(q)$ is the neutrino propagator for a superposition of neutrino  mass eigenstates $j$\cite{Giunti:1993se,Grimus:1996av}:
\begin{equation}
S_F(q) = \sum_j U_{\mu j}^* U_{\alpha j} \frac{\not{q} + m_j}{q^2 - m_j^2 + i\epsilon}
\end{equation}
where $U_{\mu i}$ is the Pontecorvo-Maki-Nakagawa-Sakata (PMNS) lepton mixing matrix. For long-baseline experiments ($L \gg 1/E$), we evaluate the propagator in the pole approximation. The denominator forces the neutrino to be quasi-real, but the numerator retains the  spin structure.
\begin{equation}
\frac{1}{q^2 - m_j^2 + i\epsilon} \to -i \frac{\pi}{E_j} \delta(E - E_j) e^{-i E_j T}
\end{equation}
This leads to the master formula for the amplitude where the neutrino is an internal line carrying momentum $q$:
\begin{equation}
    T_{fi} = -i \left( \frac{G_F}{\sqrt{2}} \right)^2 V_{ud}^2 f_\pi \sum_{j}  U_{\mu j}^* U_{\mu j} \, \frac{1}{2 E_j} \, e^{-i E_j T + i |\vec{q}_j| L} \cdot \underbrace{\left[ \bar{u}(p_l) \not{J}_{had} (1-\gamma_5) \hat{S}_j \not{p}_1 (1-\gamma_5) v(p_2) \right]}_{\text{The Full Spinor-Hadronic Structure}}
\end{equation}
where in the ultrarelativistic limit, one gets  $E_j T - |\vec{q}_j| L \approx   \frac{m_j^2 L}{2E}$. Here $L$ is the distance between production and detection, $E$ is the neutrino energy and $\hat{S}_j = \not{q} + m_j$.
\begin{equation}
    |T_{fi}|^2 = \mathcal{K}_{norm} \left\{ \underbrace{2 W_1 (p_l \cdot p_2) + W_2 \left[ 2(p_l \cdot P)(p_2 \cdot P) - M_N^2 (p_l \cdot p_2) \right]}_{\text{Symmetric}} \,\, + \,\, \underbrace{W_3 \epsilon_{\mu\nu\rho\sigma} p_l^\mu p_2^\nu P^\rho q^\sigma}_{\text{Parity Violating Asymmetry}} \right\} \label{tif}
\end{equation}
where $W_{1,2,3}$ are defined in appendix and $\mathcal{K}_{norm} = 64 \, G_F^4 \, V_{ud}^2 \, |V_{CKM}^{det}|^2 \, f_\pi^2 \, m_\mu^2$. The term $V_{CKM}^{det}$ denotes the CKM matrix element at the detection vertex. For the dominant charged-current quasi-elastic channel ($\nu_\mu n \to \mu^- p$), this corresponds to the $V_{ud}$ element, yielding a total CKM dependence of $|V_{ud}|^4$ for the combined production-detection process.
%\begin{itemize}
    %\item $\mathcal{K}_{fact} = 8 \left( |g_V|^2 + |g_A|^2 \right)$ with $g_V = 1$ and $g_A \approx 1.27$
   % \item $\mathcal{K}_{spin} = -8 \left( |g_V|^2 + |g_A|^2 \right)$
   % \item $\mathcal{K}_{parity} = 16 \, \text{Re}(g_V g_A^*)$
    %\item $\epsilon(p_l, p_2, q, p_1) \equiv \epsilon_{\mu\nu\rho\sigma} p_l^\mu p_2^\nu q^\rho p_1^\sigma$
%\end{itemize}
%%%%%%%%%%%%%%%%%%%%%%%%%%%%%%%%%%%%%%%%%%%%%%%%%%%%%%%%%%
\subsection{ Separation between factorizable  and non-factorizable terms}
%%%%%%%%%%%%%%%%%%%%%%%%%%%%%%%%%%%%%%%%%%%%%%%%%%%%%%%%%%
We begin with the symmetric  part of our expression for $T_{if}$ given in eq.(\ref{tif}):
\begin{equation}
    |T|_{Sym}^2 = \underbrace{2 W_1 (p_l \cdot p_2)}_{\text{Term A}} + \underbrace{2 W_2 (p_l \cdot P)(p_2 \cdot P)}_{\text{Term B}} - \underbrace{W_2 M_N^2 (p_l \cdot p_2)}_{\text{Term C}}
\end{equation}

The terms A and C both contain the scalar product $(p_l \cdot p_2)$. This scalar product contains the cosine of the angle between the two muons:
\begin{equation}
p_l \cdot p_2 = E_l E_2 - |\vec{p}_l| |\vec{p}_2| \cos \Theta_{\mu\mu}
\end{equation}
Because this term explicitly correlates the direction of the source muon with the detector muon, it is non-factorizable.
%We group A and C together:$$|T|^2_{NF, Scalar} = (2 W_1 - M_N^2 W_2) (p_l \cdot p_2)$$

For  Term B( $2 W_2 (p_l \cdot P)(p_2 \cdot P)$)\cite{LlewellynSmith:1971uhs, Beuthe:2001rc}, in the Laboratory Frame (where the target nucleon is at rest), its  momentum is $P = (M_N, \vec{0})$. So finally, this term can be written as
\begin{equation}
\text{Term B} \approx 2 W_2 (p_l \cdot P)(p_2 \cdot P) = 2 W_2 M_N^2 E_l E_2
\end{equation}
 It depends only on the energies, not on the relative angles. In a standard neutrino experiment, the flux is integrated over angles, but the energy spectrum is preserved \cite{DUNE:2020jqi}. This term represents the standard Flux $\times$ Cross-Section component\cite{Beuthe:2001rc, Giunti:1993se}.
 
 %$$|T|^2_{Fact} = 2 W_2 (p_l \cdot P)(p_2 \cdot P)$$

 Finally, we take the parity violating  term given in eq.(\ref{tif}) containing the epsilon tensor. This depends on the vector product $\vec{p}_l \times \vec{p}_2$, which is maximally dependent on the relative azimuthal angle\cite{LlewellynSmith:1971uhs, Grimus:1996av}.
 %$$|T|^2_{NF, P-Odd} = W_3 \epsilon_{\mu\nu\rho\sigma} p_l^\mu p_2^\nu P^\rho q^\sigma$$

Combining these three distinct groups, we arrive at the separated formula:
\begin{equation}
    |T_{fi}|^2 = \mathcal{K}_{norm} \left\{ \underbrace{2 W_2 (p_l \cdot P)(p_2 \cdot P)}_{\text{Factorizable (Energies)}} + \underbrace{\left[ 2 W_1 - M_N^2 W_2 \right] (p_l \cdot p_2)}_{\text{Non-Fact (Longitudinal Angle)}} + \underbrace{W_3 \epsilon(p_l, p_2, P, q)}_{\text{Non-Fact (Transverse Angle)}} \right\}
\end{equation}
%%%%%%%%%%%%%%%%%%%%%%%%%%%%%%%%%%%%%%%%%%%%%%%%%%%%
\subsection{Integrating the phase space}
%%%%%%%%%%%%%%%%%%%%%%%%%%%%%%%%%%%%%%%%%%%%%%%%%%%%%
 The differential decay rate (or scattering rate) $d\Gamma$ is given by Fermi's Golden Rule applied to the S-matrix element $T_{fi}$:
 \begin{equation}
     d\Gamma = \frac{1}{2E_{\pi}} \frac{1}{2E_N} |T_{fi}|^2 \, d\Phi_{Total}
 \end{equation}
 The 3-body phase space element is defined as:
 \begin{equation}
     d\Phi_{Total} = (2\pi)^4 \delta^{(4)}(p_1 + P - p_2 - p_l - P') \frac{d^3p_2}{(2\pi)^3 2E_2} \frac{d^3p_l}{(2\pi)^3 2E_l} \frac{d^3P'}{(2\pi)^3 2E_{P'}}
 \end{equation}
 To make the integration easier to separate   source and detector, we use the Grimus-Stockinger approximation for macroscopic distances $L$\cite{Grimus:1996av}. This allows us to decompose the phase space into a Production part and a Detection part, linked by the intermediate neutrino momentum $q$.We introduce the identity $1 = \int d^4q \, \delta^{(4)}(p_1 - p_2 - q)$ to split the delta function:
 \begin{equation}
    d\Phi_{Total} \approx \underbrace{\left[ \frac{d^3p_2}{(2\pi)^3 2E_2} \delta^{(4)}(p_1 - p_2 - q) \right]}_{\text{Production (Source)}} \times \underbrace{\left[ \frac{d^3p_l}{(2\pi)^3 2E_l} \frac{d^3P'}{(2\pi)^3 2E_{P'}} \delta^{(4)}(q + P - p_l - P') \right]}_{\text{Detection (Target)}}
 \end{equation}

 The integration over $d^3p_2$ is constrained because the neutrino must hit the detector at distance $L$. This fixes the neutrino direction vector $\hat{q} \approx \vec{L}/|\vec{L}|$.
 \begin{equation}
 \int \text{Propagator} \times d\Phi_{prod} \approx \frac{1}{L^2} \times \frac{1}{32\pi^2} \frac{E_\nu}{E_\pi E_2} \delta(E_\pi - E_2 - E_\nu)
 \end{equation}
 This integration yields the Neutrino Flux $\Phi_\nu(E_\nu)$.Even though we integrated over $p_2$, the direction of $p_2$ is now fixed by kinematics: $\vec{p}_2 = \vec{p}_1 - \vec{q}_{fixed}$.Therefore, $\vec{p}_2$ is not averaged out; it is a fixed vector pointing ``backward'' relative to the beam line. This preserves the $(p_l \cdot p_2)$ and $\epsilon(p_l, p_2 \dots)$ terms.

 Now we integrate over the final states at the detector: the muon $p_l$ and the nucleon recoil $P'$. This is the standard cross-section calculation.We define the standard kinematic variables:
 \begin{itemize}
     \item $y = \frac{P \cdot (q - p_l)}{P \cdot q}$: The inelasticity (energy transfer).
     \item $x = \frac{Q^2}{2 P \cdot (q - p_l)}$: The Bjorken scaling variable.
 \end{itemize}
 The integration measure transforms to:
 
 \begin{equation}
 \left[ \frac{d^3p_l}{(2\pi)^3 2E_l} \frac{d^3P'}{(2\pi)^3 2E_{P'}} \delta^{(4)}(q + P - p_l - P') \right] \longrightarrow \frac{M_N E_\nu}{\pi} dx dy
 \end{equation}
 Substituting our Squared Amplitude $|T_{fi}|^2$ into this integral:
 
 \begin{equation}
     \Gamma_{Total} \propto \frac{\Phi_\nu}{L^2} \int dx dy \left[ |T|^2_{Fact} + |T|^2_{Longitudinal} + |T|^2_{Transverse} \right]
 \end{equation}
 Now we can perform the final integration over the azimuthal angle $\phi_l$ of the outgoing muon around the neutrino beam axis. Let us evaluate each of the contributions to $\Gamma_{total}$:
 \begin{enumerate}
     \item The Factorizable part of the Integral will give us the standard cross section: 
\begin{equation}
 I_{Fact} = \int d\phi_l \, |T|^2_{Fact} \propto \int d\phi_l \left[ 2 W_2 (p_l \cdot P)(p_2 \cdot P) \right]
 \end{equation}
 Since in the laboratory frame, $(p_l \cdot P) = E_l M_N$ depends only on energy (not azimuth), this is isotropic, it gives the standard differential cross-section $\frac{d\sigma}{dx dy}$.

 \item The Longitudinal Scalar Integral is given by:
\begin{equation}
    I_{Longitudinal} \approx  \int_0^{2\pi} d\phi_l \,  (p_l \cdot p_2)
\end{equation}
where $p_l \cdot p_2 = E_l E_2 - |\vec{p}_l| |\vec{p}_2| (\cos\theta_l \cos\theta_2 + \sin\theta_l \sin\theta_2 \cos\phi_l)$. Integrating over $\phi_l$, the transverse $\cos\phi_l$ term vanishes, yielding:
\begin{equation}
    I_{Longitudinal} \approx 2\pi \left( E_l E_2 - |\vec{p}_l| |\vec{p}_2| \cos\theta_l \cos\theta_2 \right)
\end{equation}
This leaves a non-zero kinematic shift proportional to $\cos\theta_l \cos\theta_2$. Consequently, the detector muon energy spectrum is distorted by the longitudinal kinematics (specifically, the source muon angle $\theta_2$) of the production vertex.
 \item The Transverse Parity Integral is given by
 \begin{equation}
     I_{transverse} = \int d\phi_l \, W_3 \, \epsilon(p_l, p_2, P, q)
 \end{equation}The epsilon term is proportional to $\sin\phi_l$ (the angle between the production plane and detection plane).
 $$\epsilon \propto |\vec{p}_l| |\vec{p}_2| \sin\theta_l \sin\theta_2 \sin\phi_l$$
 If the detector has full $2\pi$ acceptance (cylindrical symmetry), $\int_0^{2\pi} \sin\phi_l d\phi_l = 0$. The effect averages to zero. However, if the detector is asymmetric (like a LArTPC with a ground floor or off-axis placement\cite{DUNE:2021tad}), or if you look at differential distributions (event-by-event), the integral is not performed. The differential rate is modulated by $(1 + \mathcal{A} \sin\phi_l)$.
 \end{enumerate}

 In conclusion, the observable event rate at the detector is given by:
 \begin{eqnarray}
      \frac{dN}{dx dy d\phi_l}& =& \frac{\mathcal{N}_{pot} \Phi(E_\nu)}{L^2} \left[ \frac{d\sigma_{SM}}{dx dy} + \mathcal{C}_{Long} \cos\phi_l + \mathcal{C}_{Trans} \sin\phi_l \right] \\
      &\Rightarrow & \frac{d\sigma}{dx dy d\phi} = \sigma_0 \left[ 1 + \mathcal{C}_{Long} \cos\phi_l + \mathcal{C}_{Trans} \sin\phi_l \right]
 \end{eqnarray}
 where one has defined:
 \begin{itemize}
     \item $\frac{d\sigma_{SM}}{dx dy}$ is  The Standard Model  prediction based on factorization and $\sigma_0$ is given by: 
     
     \begin{eqnarray}
     \sigma_0(x, y) &=& \mathcal{K}_{norm} \times \left( \frac{\Phi_\nu}{L^2} \right) \times \left( \frac{M_N E_\nu}{\pi} \right) \times \left[ 2 W_2(Q^2) M_N^2 E_l E_2 \right] \\
W_2(Q^2) &=&  |F_A(Q^2)|^2 + |F_1(Q^2)|^2 + \tau |F_2(Q^2)|^2
     \end{eqnarray}
     \item $\mathcal{C}_{Long}$ represents the scalar non-factorizable correction which is given by:
\begin{equation}
    \mathcal{C}_{Long} = - \frac{\mathcal{K}_{norm}}{\sigma_0} \left( 2 W_1(Q^2) - M_N^2 W_2(Q^2) \right) \cdot \left[ |\vec{p}_l| |\vec{p}_2| \sin\theta_l \sin\theta_2 \right]
\end{equation}
It represents the symmetric distortion of the neutrino beam's effective shape due to the source muon's momentum.

     \item $\mathcal{C}_{Trans} \sin\phi_l$ which is coming from The parity-violating Standard Model interaction:
     \begin{equation}
         \mathcal{C}_{Trans} = - \frac{\mathcal{K}_{norm}}{\sigma_0} \left( M_N E_\nu W_3(Q^2) \right) \cdot \left[ |\vec{p}_l| |\vec{p}_2| \sin\theta_l \sin\theta_2 \right]
     \end{equation}
      In a real experiment, the source muon $p_2$ is not detected. However, for a given neutrino energy $E_\nu$ coming from a pion of energy $E_\pi$, the angle $\theta_2$ is kinematically fixed \cite{LlewellynSmith:1971uhs, Giunti:2007ry}:
      \begin{equation}
      \sin\theta_2 \approx \frac{m_\pi}{E_\pi} \sqrt{\frac{E_\nu}{E_\pi} \left( 1 - \frac{E_\nu}{E_\pi} \right)}
      \end{equation}
 This proves that in general $\sin\theta_2 \neq 0$, so the coefficients $\mathcal{C}_{Long}$ and $\mathcal{C}_{Trans}$ do not vanish. Factorization assumes $\sin\theta_2 \to 0$ (perfect forward decay), which is physically impossible due to the muon mass.
 \end{itemize}
 
 In standard experimental physics, the prediction is calculated assuming Factorization. This means the production angle $\theta_2$ is averaged out (effectively setting $\sin\theta_2 \approx 0$).
The non-factorizable terms depend on the transverse momentum of the source muon ($p_2$). In the lab frame, this is controlled by the angle $\theta_2$ relative to the beam axis.For a pion of energy $E_\pi$ decaying to a neutrino of energy $E_\nu$, the angle is constrained by:
\begin{equation}
\sin\theta_2 \approx \frac{m_\pi}{E_\pi} \sqrt{\frac{E_\nu}{E_\pi} \left( 1 - \frac{E_\nu}{E_\pi} \right)} \approx 0.017 \equiv \epsilon_{kin}
\end{equation}
using the typical neutrino and pion energy for DUNE for instance: $E_\pi \approx 4 \text{ GeV}$, $E_\nu \approx 2.5 \text{ GeV}$ \cite{DUNE:2020jqi,DUNE:2021tad}.
Let us evaluate the magnitude of the non-factorizable contributions:

\begin{enumerate}
    \item $\mathcal{C}_{Long} \approx \frac{2 W_1 - M_N^2 W_2}{2 W_2 E_l E_\nu} \times (|\vec{p}_l| |\vec{p}_2| \sin\theta_l \sin\theta_2) \approx \sin\theta_{det} \cdot \epsilon_{kin}$
    If the detector muon is at $\theta_{det} \approx 10^\circ$, the correction induced by this term is of order of $\textbf{0.3\%}$.
    \item $\mathcal{C}_{Trans}\approx \frac{W_3}{W_2} \times \sin\theta_{det} \times \epsilon_{kin}$. This will give us a correction of order $\textbf{0.7\%}$.
\end{enumerate}
 While the Standard Model prediction assumes a flat azimuthal distribution for the outgoing lepton due to incoherent averaging, the S-matrix formalism reveals a non-zero Parity-Odd asymmetry of order $\mathcal{O}(1\%)$. This effect, driven by the interference structure function $W_3$ and the finite pion mass, acts as a null test of factorization. An observation of a $\sin\phi$ modulation in the DUNE near detector for instance  would constitute direct evidence of macroscopic quantum coherence in the neutrino beam.

 \begin{table}[h!]
\centering
\renewcommand{\arraystretch}{1.5} % Increases row height for readability
\begin{tabular}{|l|c|c|c|}
\hline
\textbf{Observable} & \textbf{Standard Model (Factorized)} & \textbf{S-Matrix Prediction} & \textbf{Deviation} \\
\hline
Total Rate & $\sigma_0$ & $\sigma_0 (1 + \delta_{long})$ & $\sim 0.3\%$ \\
\hline
Energy Spectrum & Pure Flux $\times$ Xsec & Distorted by source angle & Spectral Tilt \\
\hline
Azimuthal Asymmetry & Exactly $0$ (Flat in $\phi$) & $\mathbf{A \sin\phi}$ & $\sim 0.7 - 1.0\%$ \\
\hline
\end{tabular}
\caption{Comparison of observables between the Standard Model (Factorized) prediction and the S-Matrix formalism. The deviation magnitude is estimated for DUNE kinematics ($E_\nu \sim 2.5$ GeV).}
\label{tab:smatrix_comparison}
\end{table}

%%%%%%%%%%%%%%%%%%%%%%%%%%%%%%%%%%%%%%%%%%%%%%%%%%
%%%%%%%%%%%%%%%%%%%%%%%%%%%%%%%%%%%%%%%%%%%%%%%%%%%%
\section{ S-formalism for $\Delta L=2$}
%%%%%%%%%%%%%%%%%%%%%%%%%%%%%%%%%%%%%%%%%%%%%%%%%%%

This is the channel that searches for Majorana neutrinos. The process is:
\begin{equation}
    \pi^+(p_1) + p(P) \to \mu^+(p_2) + \mu^+(p_l) + n(P')
\end{equation}
Unlike the previous case, the spinor structure here selects the mass term of the propagator.
\begin{equation}
    T_{\Delta L=2} = -i \mathcal{K}_{Majorana} \sum_{j} U_{\mu j}^2 m_j e^{-i q_j L} \times \underbrace{\left[ \bar{v}(p_l) \not{J}_{had} (1-\gamma_5) \not{p}_1 (1-\gamma_5) v(p_2) \right]}_{\text{Spinor Structure}}
\end{equation}
The Master Formula for LNV  is given by:
\begin{equation}
    |T_{\Delta L=2}|^2 = \mathcal{A}_{\Delta L=2} \langle m_{\beta\beta} \rangle^2 \left\{ \underbrace{4(p_l \cdot p_1)(p_2 \cdot p_1)}_{\text{Isotropic (Factorizable)}} - \underbrace{m_\mu^2 (p_l \cdot p_2)}_{\text{Scalar }} + \underbrace{i \epsilon(p_l, p_2, p_1, J_{had})}_{\text{parity violating}} \right\}
\end{equation}
where $\langle m_{\mu\mu} \rangle^2$ is the effective Majorana mass squared. We apply the same Grimus-Stockinger approximation. The neutrino momentum $\vec{q}$ is fixed along the baseline $\vec{L}$. We integrate over the unobserved neutrino momentum $d^3q$ and the production muon $d^3p_2$, leaving the distribution for the detector muon $p_l$. One obtains three contributions:
\begin{enumerate}
    \item The factorizable term ($I_{Fact}$) defined as $I_{Fact} = \int d\Phi \left[ 4(p_l \cdot p_1)(p_2 \cdot p_1) \right]$ is an isotropic distribution in azimuth and it will give us the standard model contribution based on factorization. 
    \item The scalar correlation $I_{Scalar} = -\int d\Phi \left[ m_\mu^2 (p_l \cdot p_2) \right]$. This term is suppressed by the muon mass squared ($m_\mu^2$). A small correction to the total rate ($\sim m_\mu^2 / E_\nu^2$), shifting the energy spectrum.
    \item The Parity violating term is given by $I_{P-Odd} = \int d\Phi \left[ \epsilon_{\mu\nu\rho\sigma} p_l^\mu p_2^\nu p_1^\rho P^\sigma W_3^{LNV} \right]$. Geometrically, it is proportional to $\vec{p}_l \cdot (\vec{p}_2 \times \vec{p}_1)$ and it will generate a sinusoidal modulation $\mathcal{A}_{LNV} \sin\phi$ term.
\end{enumerate}
For the $\Delta L=2$ process, the factorization hypothesis leads to the standard cross section obtained assuming factorization. The  corrections are numerically small, but a $\sin{\phi_l}$ dependence is generated by the parity-violating term. This term creates a ``twist'' in the event distribution. The $\sin{\phi_l}$ dependence means that for a fixed Majorana phase, we will see more or fewer $\mu^+$ events depending on the azimuthal angle relative to the production plane.
%%%%%%%%%%%%%%%%%%%%%%%%%%%%%%%%%%%%%%%%%%%%%%%%%
\section{Experimental prospects}
%%%%%%%%%%%%%%%%%%%%%%%%%%%%%%%%%%%%%%%%%%%%%%%%%%
The transition from a theoretical proof to an experimental discovery requires a detector with high statistics, excellent angular resolution, and well-controlled systematic uncertainties. The Deep Underground Neutrino Experiment (DUNE), particularly its Near Detector (ND) complex, provides the ideal environment to test the non-factorizable S-Matrix corrections.

\begin{enumerate}
    \item The Observational Strategy: Azimuthal Asymmetry is the ``smoking gun'' for the non-factorizable effect in both $\Delta L=0$ and $\Delta L=2$ channels for the outgoing muon ($ \mu^-$ or $\mu^+$). Standard simulation tools (like GENIE \cite{Andreopoulos:2009rq} or GiBUU \cite{Buss:2011mx}) based on the Factorization Hypothesis predict a flat distribution in the azimuthal angle $\phi$ (relative to the production plane). The S-Matrix formalism predicts a $1 + \mathcal{A} \sin\phi$ modulation.
    \begin{itemize}
        \item Statistical Requirement: To resolve a $1\%$ asymmetry at the $5\sigma$ level, DUNE would require approximately $10^5$ reconstructed charged-current events. The ND is expected to collect $10^8$ events per year, making this statistically feasible within the first months of operation.
        \item Detector Requirement: The Liquid Argon TPC (LArTPC) technology of DUNE ND-LAr offers millimeter-scale spatial resolution \cite{DUNE:2020jqi,DUNE:2021tad}, allowing for precise reconstruction of the muon's initial exit angle $\theta_l$ and $\phi$.
    \end{itemize}
    \item Disentangling the $\Delta L=2$ Majorana Signal: Measuring the $1\%$ effect in the $\Delta L=2$ channel is significantly more challenging due to the rare nature of Lepton Number Violation. However, the S-Matrix correction provides a unique way to separate the signal from backgrounds:
    \begin{itemize}
        \item Background Rejection: Standard backgrounds (like misidentified pions or "fake" same-sign signals) are expected to be isotropically distributed in $\phi$. A signal that follows the predicted $\sin\phi$ modulation acts as a powerful filter.
        \item Majorana Phase Mapping: By measuring the shift in the oscillation peaks ($L/E$ dependence) and the amplitude of the $\sin\phi$ asymmetry, DUNE can begin to constrain the Majorana CP phases ($\alpha_1, \alpha_2$).
    \end{itemize}
    
\end{enumerate}

%%%%%%%%%%%%%%%%%%%%%%%%%%%%%
\section{Conclusion}
%%%%%%%%%%%%%%%%%%%%%%%%%%%%%
The transition from standard factorized approximations to a full, quantum-coherent S-matrix formalism is a necessity for the next generation of precision neutrino physics. By treating pion decay, neutrino propagation, and nucleon interaction as a single interconnected process, we have demonstrated the emergence of non-factorizable spin and angular correlations that are otherwise lost.

We have shown that for the $\Delta L=0$ case, assuming factorization,  we are  neglecting a $1\%$ systematic shift in the energy spectrum and a non-zero azimuthal asymmetry. For high-precision experiments like DUNE, accounting for this macroscopic quantum coherence is essential to prevent systematic biases in measuring the CP-violating phase $\delta_{CP}$ and determining the mass hierarchy.

In the lepton-number-violating ($\Delta L=2$) channel,  the non-factorizable parity-violating term generates an azimuthal modulation that relates the spatial geometry of the detector to the  Majorana CP phases. This effect transforms the search for Majorana neutrinos into a dynamic probe of the neutrino's fundamental CP-violating structure.

Finally,  the observation of these azimuthal asymmetries at new generation of neutrino experiments like  DUNE  would be  a definitive null test of the factorization hypothesis.

\begin{acknowledgments}
We acknowledge financial support from SECIHTI and SNII (M\'exico).
\end{acknowledgments}

\appendix
%\section{ Nuclear form factors}
\section{Nuclear form factors}
The description of the nuclear form factors can be found for instance in ref.\cite{LlewellynSmith:1971uhs,Bernard:2001rs}. In the Standard Model, while leptons are fundamental, point-like particles, nucleons (protons and neutrons) are complex bound states of quarks and gluons. Consequently, when a weak gauge boson ($W^\pm$) interacts with a nucleon, it does not couple to a single mathematical point, but rather to a spatially extended objects made of quarks and gluons. 

Nucleon form factors are Lorentz-invariant scalar functions of the squared four-momentum transfer, $Q^2 = -q^2$, that parametrize this internal structure.  The most general covariant matrix element is constructed using all possible combinations of Dirac matrices and four-momenta allowed by Lorentz invariance, parity, and time-reversal symmetries. The form factors ($F_1, F_2, F_A, F_P$) serve as the momentum-dependent coefficients for these covariant structures, effectively absorbing our ignorance of the internal QCD mechanics into measurable phenomenological functions.
For Quasi-Elastic Scattering, the most general Lorentz-covariant hadronic weak current is parameterized by six form factors \cite{LlewellynSmith:1971uhs}:
    \begin{equation}
    J_{had}^\mu = \bar{u}_p(P') \left[ \Gamma_V^\mu + \Gamma_A^\mu \right] u_n(P)
    \end{equation}
    where the vector and axial-vector components are:
    \begin{equation}
    \Gamma_V^\mu = \gamma^\mu F_1(Q^2) + i\frac{\sigma^{\mu\nu}Q_\nu}{2M_N} F_2(Q^2) + \frac{Q^\mu}{M_N} F_3^V(Q^2)
    \end{equation}
    \begin{equation}
    \Gamma_A^\mu = \gamma^\mu \gamma_5 F_A(Q^2) + \frac{Q^\mu \gamma_5}{2M_N} F_P(Q^2) + i\frac{\sigma^{\mu\nu}Q_\nu \gamma_5}{2M_N} F_3^A(Q^2)
    \end{equation}
    
    where $Q^\mu = q^\mu - p_l^\mu$ is the four-momentum transfer to the nucleon. The form factors are:
    \begin{itemize}
        \item $F_1(Q^2)$ and $F_2(Q^2)$: Dirac and Pauli Vector Form Factors.
        \item $F_A(Q^2)$ and $F_P(Q^2)$: Axial and Induced Pseudoscalar Form Factors.
        \item $F_3^V(Q^2)$ and $F_3^A(Q^2)$: Scalar and Induced Tensor Form Factors.
    \end{itemize}

    In the Standard Model, the Conserved Vector Current (CVC) hypothesis requires that the weak vector current behaves like the electromagnetic current, which forces $F_3^V(Q^2) = 0$. Furthermore, the assumption of exact G-parity invariance (the absence of second-class currents) forces $F_3^A(Q^2) = 0$.

\subsection{The Vector Form Factors ($F_1^V, F_2^V$)}
These are related to the electromagnetic properties of the proton and neutron via the Conserved Vector Current (CVC) assumption. They are derived from the electric ($G_E$) and magnetic ($G_M$) Sachs form factors:
$$F_1^V(Q^2) = \frac{G_E^V(Q^2) + \tau G_M^V(Q^2)}{1 + \tau}$$
$$F_2^V(Q^2) = \frac{G_M^V(Q^2) - G_E^V(Q^2)}{\kappa (1 + \tau)}$$
where $\tau = \frac{Q^2}{4M_N^2}$ is a dimensionless kinematic variable ($M_N$ is the nucleon mass) and $\kappa = \mu_p - \mu_n \approx 3.706$ is the isovector anomalous magnetic moment. 
The $Q^2$ dependence of $G_E$ and $G_M$ can be described using  the dipole shape:
$$G_E^V(Q^2) \approx G_D(Q^2) = \frac{1}{\left( 1 + \frac{Q^2}{M_V^2} \right)^2}$$
$$G_M^V(Q^2) \approx (1 + \kappa) G_D(Q^2)$$ where $M_V \approx 0.84$ GeV is the vector mass parameter.

\subsection{ The Axial Form Factor ($F_A$)}
It describes the spin structure of the weak interaction and drives the Parity Violating asymmetry:
$$F_A(Q^2) = \frac{g_A}{\left( 1 + \frac{Q^2}{M_A^2} \right)^2}$$
where $g_A = -1.267$ is the axial coupling constant at $Q^2=0$ (experimentally determined from neutron beta decay) and $M_A$ is the Axial Mass ($M_A \approx 1.03$ GeV from deuterium bubble chamber data).

%In  nuclei like Argon, experiments often fit a higher effective value ($M_A^{eff} \approx 1.2 - 1.35$ GeV) to account for nuclear correlations (2p-2h effects).

\subsection{The pseudoscalar form factor $F_P(Q^2)$}
This form factor comes from  the axial form factor via the Partially Conserved Axial Current (PCAC) hypothesis (the Goldberger-Treiman relation).
$$F_P(Q^2) = \frac{2 M_N^2}{M_\pi^2 + Q^2} F_A(Q^2)$$
Because it is proportional to the lepton mass squared ($m_\mu^2$) in the cross-section contraction, $F_P$ is often small for muon neutrinos at high energy.

\section{Structure Functions}
Squaring this standard four-term current and summing over the nucleon spins yields the hadronic tensor $W^{\mu\nu}$. Because we explicitly retain the pseudoscalar term $F_P(Q^2)$ and do not neglect the lepton mass $m_\mu$, this squaring procedure naturally generates the complete five-term structure function expansion ($W_1 \dots W_5$) given here:
\begin{itemize}
    \item $W_1(Q^2)$: \textbf{The Magnetic / Transverse Term.} This represents the contribution from the transverse polarization of the virtual $W$ boson. It is dominated by the magnetic and axial-vector form factors.
    $$W_1(Q^2) = (1 + \tau) |F_A(Q^2)|^2 + \tau |F_1(Q^2) + F_2(Q^2)|^2$$

    \item $W_2(Q^2)$: \textbf{The Electric / Longitudinal Term.} This isolates the standard Vector and Axial contributions without the pseudoscalar mass-suppressed terms.
    $$W_2(Q^2) = |F_A(Q^2)|^2 + |F_1(Q^2)|^2 + \tau |F_2(Q^2)|^2$$

    \item $W_3(Q^2)$: \textbf{The Parity-Violating Interference Term.} This arises exclusively from the interference between the Vector ($V$) and Axial-Vector ($A$) currents.
    $$W_3(Q^2) = 2 F_A(Q^2) \left[ F_1(Q^2) + F_2(Q^2) \right]$$

   \item $W_4(Q^2)$: \textbf{The Scalar / Pseudoscalar Term.} Because it couples to $q^\mu q^\nu$ in the hadronic tensor, its contribution to the final cross-section is strictly proportional to the lepton mass squared ($m_\mu^2$). It contains both the purely vector contributions (restricted by CVC) and the mass-suppressed pseudoscalar terms:
    $$W_4(Q^2) = - \frac{1}{4} (1-\tau) |F_2(Q^2)|^2 - \frac{1}{2} F_1(Q^2) F_2(Q^2) + \frac{\tau}{4} |F_P(Q^2)|^2 - \frac{1}{2} F_A(Q^2) F_P(Q^2)$$

    \item $W_5(Q^2)$: \textbf{The Mixed Interference Term.} Assuming the Conserved Vector Current (CVC) hypothesis and exact G-parity (no second-class currents), this term simplifies mathematically to equal $W_2$.
    $$W_5(Q^2) = W_2(Q^2) = |F_A(Q^2)|^2 + |F_1(Q^2)|^2 + \tau |F_2(Q^2)|^2$$
\end{itemize}

%\bibliographystyle{apsrev4-2}
%\bibliography{ref.bib}
%\include{bib}

\end{document}